\documentclass[11pt,a4paper]{article}
\usepackage{amsfonts}
\usepackage{amsmath,amssymb}
\usepackage{epsfig,graphicx}

\input{tcilatex}

\begin{document}

\begin{center}
{\huge Photon and photino as Nambu-Goldstone zero modes in an emergent SUSY
QED }

\bigskip \bigskip\bigskip

\textbf{J.L.~Chkareuli}$^{1,2}$

\bigskip

$^{1}$\textit{Center for Elementary Particle Physics, Ilia State University,
0162 Tbilisi, Georgia\ \vspace{0pt}\\[0pt]
} $^{2}$\textit{E. Andronikashvili} \textit{Institute of Physics, 0177
Tbilisi, Georgia\ } \bigskip

\bigskip

\bigskip

\bigskip

\bigskip

\bigskip

\bigskip

\textbf{Abstract} \bigskip \bigskip
\end{center}

We argue that supersymmetry with its well known advantages, such as
naturalness, grand unification and dark matter candidate seems to possess
one more attractive feature: it may trigger, through its own spontaneous
violation in the visible sector, a dynamical generation of gauge fields as
massless Nambu-Goldstone modes during which physical Lorentz invariance
itself is ultimately preserved. We consider the supersymmetric QED model
extended by an arbitrary polynomial potential of massive vector superfield
that breaks gauge invariance in the SUSY invariant phase. However, the
requirement of vacuum stability in such class of models makes both
supersymmetry and Lorentz invariance to become spontaneously broken. As a
consequence, massless photino and photon appear as the corresponding
Nambu-Goldstone zero modes in an emergent SUSY QED, and also a special gauge
invariance is simultaneously generated. Due to this invariance all
observable relativistically noninvariant effects appear to be completely
cancelled out among themselves and physical Lorentz invariance is recovered.
Nevertheless, such theories may have an inevitable observational evidence in
terms of the goldstino-photino like state presented in the low-energy
particle spectrum. Its study is of a special interest for this class of SUSY
models that, apart from some indication of an emergence nature of QED and
the Standard Model, may appreciably extend the scope of SUSY breaking
physics being actively studied in recent years.

\bigskip

\bigskip

\bigskip

\bigskip

\bigskip

\bigskip

\bigskip

\thispagestyle{empty}\newpage

\section{Introduction and overview}

It is long believed that spontaneous Lorentz invariance violation (SLIV) may
lead to an emergence of massless Nambu-Goldstone (NG) zero modes \cite{NJL}
which are identified with photons and other gauge fields appearing in the
Standard Model. This old idea \cite{bjorken} supported by a close analogy
with the dynamical origin of massless particle excitations for spontaneously
broken internal symmetries has gained new impetus in recent years. On the
other hand, besides its generic implication to a possible origin of physical
gauge fields \cite{cfn,jb,kraus,jen,bluhm} in a conventional quantum field
theory (QFT) framework, there are many different contexts in literature
where Lorentz violation may stem in itself from string theory \cite{alan},
quantum gravity \cite{ted} or any unspecified dynamics at an ultraviolet
scale perhaps related to the Planck scale \cite{refs,2,3}. Though we are
mainly related to the spontaneous Lorentz violation in QFT, particularly in
QED and Standard Model, we give below some brief comments on other
approaches as well to make clearer the aims and results of the present work.

\subsection{Vector NG bosons in gauge theories. Inactive SLIV}

When speaking about SLIV, one important thing to notice is that, in contrast
to the spontaneous violation of internal symmetries, it seems not to
necessarily imply a physical breakdown of Lorentz invariance. Rather, when
appearing in a gauge theory framework, this may ultimately result in a
noncovariant gauge choice in an otherwise gauge invariant and Lorentz
invariant theory. In substance, the SLIV ansatz, due to which the vector
field develops a vacuum expectation value (VEV) 
\begin{equation}
<A_{\mu }(x)>\text{ }=n_{\mu }M  \label{vev1}
\end{equation}%
(where $n_{\mu }$ is a properly-oriented unit Lorentz vector, $n^{2}=n_{\mu
}n^{\mu }=\pm 1$, while $M$ is the proposed SLIV scale), may itself be
treated as a pure gauge transformation with a gauge function linear in
coordinates, $\omega (x)=$ $n_{\mu }x^{\mu }M$. From this viewpoint gauge
invariance in QED leads to the conversion of SLIV into gauge degrees of
freedom of the massless photon emerged.

A good example for such a kind of SLIV, which we call the "inactive" SLIV
hereafter, is provided by the nonlinearly realized Lorentz symmetry for
underlying vector field $A_{\mu }(x)$ through the length-fixing constraint%
\begin{equation}
A_{\mu }A^{\mu }=n^{2}M^{2}\text{ .}  \label{const}
\end{equation}%
This constraint in the gauge invariant QED framework was first studied by
Nambu a long ago \cite{nambu}, and in more detail in recent years \cite%
{az,kep,jej,urr,gra}. The constraint (\ref{const}) is in fact very similar
to the constraint appearing in the nonlinear $\sigma $-model for pions \cite%
{GL}, $\sigma ^{2}+\pi ^{2}=f_{\pi }^{2}$, where $f_{\pi }$ is the pion
decay constant. Rather than impose by postulate, the constraint (\ref{const}%
) may be implemented into the standard QED Lagrangian extended by the
invariant Lagrange multiplier term

\begin{equation}
\mathcal{L}=L_{QED}-\frac{\lambda }{2}\left( A_{\mu }A^{\mu
}-n^{2}M^{2}\right)  \label{lag}
\end{equation}%
provided that initial values for all fields (and their momenta) involved are
chosen so as to restrict the phase space to values with a vanishing
multiplier function $\lambda (x)$, $\lambda =0$. Otherwise, as was shown in 
\cite{vru} (see also \cite{urr}), it might be problematic to have the
ghost-free QED model with a positive Hamiltonian\footnote{%
Note that this solution with the basic Lagrangian multiplier field $\lambda
(x)$ being vanished can technically be realized by introducing some
additional Lagrange multiplier term of the type $\xi \lambda ^{2}$, where $%
\xi (x)$ is a new multiplier field. One can now easily confirm that a
variation of the modified Lagrangia $\mathcal{L}+$ $\xi \lambda ^{2}$ with
respect to the $\xi $ field leads to the condition $\lambda =0$, whereas a
variation with respect to the basic multiplier field $\lambda $ preserves
the vector field constraint (\ref{const}).}.

One way or another, the constraint (\ref{const}) means in essence that the
vector field $A_{\mu }$ develops the VEV (\ref{vev1}) and Lorentz symmetry $%
SO(1,3)$ breaks down to $SO(3)$ or $SO(1,2)$ depending on whether the unit
vector $n_{\mu }$ is time-like ($n^{2}>0$) or space-like ($n^{2}<0$). The
point, however, is that, in sharp contrast to the nonlinear $\sigma $ model
for pions, the nonlinear QED theory, due to gauge invariance in the starting
Lagrangian $L_{QED}$, leaves physical Lorentz invariance intact. Indeed, the
nonlinear QED contains a plethora of Lorentz and $CPT$ violating couplings
when it is expressed in terms of the pure vector NG boson modes ($a_{\mu }$)
associated with a physical photon 
\begin{equation}
A_{\mu }=a_{\mu }+n_{\mu }(M^{2}-n^{2}a^{2})^{\frac{1}{2}}\text{ , \ }n_{\mu
}a_{\mu }=0\text{ \ \ (}a^{2}\equiv a_{\mu }a^{\mu }\text{) .}  \label{gol}
\end{equation}%
including that the effective Higgs mode given by the second term in (\ref%
{gol}) is properly expanded in a power series of $a^{2}$. However, the
contributions of all these couplings to physical processes completely cancel
out among themselves, as was shown in the tree \cite{nambu} and one-loop
approximations \cite{az}. Actually, the nonlinear constraint (\ref{const})
implemented as a supplementary condition can be interpreted in essence as a
possible gauge choice for the starting vector field $A_{\mu }$. Meanwhile
the $S$-matrix remains unaltered under such a gauge convention unless gauge
invariance in the theory turns out to be really broken (see next subsection)
rather than merely being restricted by gauge condition (\ref{const}). Later
similar result concerning the inactive SLIV in gauge theories\ was also
confirmed for spontaneously broken massive QED \cite{kep}, non-Abelian
theories \cite{jej} and tensor field gravity \cite{gra}.

Remarkably enough, the nonlinear QED model (\ref{lag}) may be considered in
some sense as being originated from a conventional QED Lagrangian extended
by the vector field potential energy terms, 
\begin{equation}
\mathcal{L}^{\prime }=L_{QED}-\frac{\boldsymbol{\lambda }}{4}\left( A_{\mu
}A^{\mu }-n^{2}M^{2}\right) ^{2}  \label{qed1}
\end{equation}%
(where $\boldsymbol{\lambda }$ is a coupling constant) rather than by the
Lagrange multiplier term. This is the simplest example of a theory being
sometimes referred to as the \textquotedblleft bumblebee\textquotedblright\
model (see \cite{bluhm} and references therein) where physical Lorentz
symmetry could in principle be spontaneously broken due to presence of an
active Higgs mode in the model. On \ the other hand, the Lagrangian (\ref%
{qed1}) taken in the limit $\boldsymbol{\lambda }\rightarrow \infty $ can
formally be regarded as the nonlinear QED. Actually, both of models are
physically equivalent in the infrared energy domain, where the Higgs mode is
considered infinitely massive. However, as was argued in \cite{vru}, a
bumblebee-like model appears generally unstable, its Hamiltonian is not
bounded from below unless the phase space sector is not limited by the
nonlinear vector field constraint $A_{\mu }A^{\mu }=n^{2}M^{2}$ (\ref{const}%
). With this condition imposed, the massive Higgs mode never appears, the
Hamiltonian is positive, and the model is physically equivalent to the
constraint-based nonlinear QED (\ref{lag}) with the inactive SLIV which does
not lead to physical Lorentz violation\footnote{%
Apart from its generic instability, the \textquotedblleft
bumblebee\textquotedblright\ model, as we will see it shortly, can not be
technically realized in a SUSY context, whereas the nonlinear QED model
successfully matches supersymmetry.}.

To summarize, we have considered above the standard QED with vector field
constraint (\ref{const}) being implemented into the Lagrangian through the
Lagrange multiplier term (\ref{lag}). In crucial contrast to an internal
symmetry breaking (say, the breaking of a chiral $SU(2)\times SU(2)$
symmetry in the nonlinear $\sigma $-model for pions) SLIV caused by a
similar $\sigma $-model type vector field constraint (\ref{const}), does not
lead to physical Lorentz violation. Indeed, though SLIV induces the vector
Goldstone-like states (\ref{gol}), all observable SLIV effects appear to be
completely canceled out among themselves due to a generic gauge invariance
of QED. We call it the inactive SLIV in the sense that one may have
Goldstone-like states in a theory but may have not a nonzero symmetry
breaking effect. This is somewhat new and unusual situation that just
happens with SLIV in gauge invariant theories (and never in an internal
symmetry breaking case). More precisely there are, in essence, two different
(though related to each other) aspects regarding the inactive SLIV. The
first is a generation of Goldstone modes which inevitably happens once the
nonlinear $\sigma $-model type constraint (\ref{const}) is put on the vector
field. The second is that gauge invariance even being restricted by this
constraint (interpreted as a gauge condition) provides a cancellation
mechanism for physical Lorentz violation. As a consequence, emergent gauge
theories induced by the inactive SLIV mechanism are in fact
indistinguishable from conventional gauge theories. Their emergent nature
can only be seen when a gauge condition is taken to be the vector field
length-fixing constraint (\ref{const}). Any other gauge, e.g. the Coulomb
gauge, is not in line with an emergent picture, since it explicitly breaks
Lorentz invariance. As to an observational evidence in favor of emergent
theories, the only way for SLIV to be activated may appear if gauge
invariance in these theories turns out to be broken in an explicit rather
than spontaneous way. As a result, the SLIV cancellation mechanism does not
work longer and one inevitably comes to physical Lorentz violation.

\subsection{Activating SLIV by gauge symmetry breaking}

Looking for some appropriate examples of physical Lorentz violation in a QFT
framework one necessarily come across a problem of proper suppression of
gauge noninvariant high-dimension couplings where such violation can in
principle occur. Remarkably enough, for QED type theories with the
supplementary vector field constraint (\ref{const}) gauge symmetry breaking
naturally appears only for five- and higher-dimensional couplings. Indeed,
all dimension-four couplings are generically gauge invariant, if the vector
field kinetic term has a standard $F_{\mu \nu }F^{\mu \nu }$ and, apart from
relativistic invariance, the restrictions related to the conservation of
parity, charge-conjugation symmetry and fermion number conservation are
generally imposed on a theory \cite{par}. With these restrictions taken, one
can easily confirm that all possible dimension-five couplings are also
combined by themselves in some would-be gauge invariant form provided that
vector field is constrained by the SLIV condition (\ref{const}). Indeed, for
charged matter fermions interacting with vector field such couplings are
generally amounted to 
\begin{equation}
L_{\dim 5}=\frac{1}{\mathcal{M}}\check{D}_{\mu }^{\ast }\overline{\psi }%
\cdot \check{D}^{\mu }\psi +\frac{G}{\mathcal{M}}A_{\mu }A^{\mu }\overline{%
\psi }\psi \text{ , \ }A_{\mu }A^{\mu }=n^{2}M^{2}\text{ .}  \label{dim5}
\end{equation}%
Such couplings could presumably become significant at an ultraviolet scale $%
\mathcal{M}$ probably being close to the Planck scale $M_{P}$. They, besides
covariant derivative terms, also include an independent "sea-gull"
fermion-vector field term with the coupling constant $G$ being in general of
the order $1$. The main point regarding the Lagrangian (\ref{dim5}) is that,
while it is gauge invariant in itself, the coupling constant $\check{e}$ in
the covariant derivative $\check{D}^{\mu }=\partial ^{\mu }+i\check{e}A^{\mu
}$ differs in general from the coupling $e$ in the covariant derivative $%
D^{\mu }=\partial ^{\mu }+ieA^{\mu }$ in the standard Dirac Lagrangian (\ref%
{lag})%
\begin{equation}
L_{QED}=-\frac{1}{4}F_{\mu \nu }F^{\mu \nu }+\overline{\psi }(i\gamma _{\mu
}D^{\mu }-m)\psi \text{ .}  \label{qed}
\end{equation}%
Therefore, gauge invariance is no longer preserved in the total Lagrangian $%
L_{QED}+$ $L_{\dim 5}$. It is worth noting that, though the high-dimension
Lagrangian part $L_{\dim 5}$ (\ref{dim5}) usually only gives some small
corrections to a conventional QED Lagrangian (\ref{qed}), the situation may
drastically change when the vector field $A_{\mu }$ develops a VEV and SLIV
occurs.

Actually, putting the SLIV parameterization (\ref{gol}) into the basic QED
Lagrangian (\ref{qed}) one comes to the truly emergent model for QED being
essentially nonlinear in the vector Goldstone modes $a_{\mu }$ associated
with photons. This model contains, among other terms, the inappropriately
large (while false, see below) Lorentz violating fermion bilinear $-eM%
\overline{\psi }(n_{\mu }\gamma ^{\mu })\psi $. This term appears when the
effective Higgs mode expansion in Goldstone modes $a_{\mu }$ (as is given in
the parametrization (\ref{gol})) is applied to the fermion current
interaction term $-e\overline{\psi }\gamma _{\mu }A^{\mu }\psi $ in the QED
Lagrangian (\ref{qed}). However, due to local invariance this bilinear term
can be gauged away by making an appropriate redefinition of the fermion
field $\psi \rightarrow e^{-ie\omega (x)}\psi $ with a gauge function $%
\omega (x)$ linear in coordinates, $\omega (x)=$ $(n_{\mu }x^{\mu })M$.
Meanwhile, the dimension-five Lagrangian $L_{\dim 5}$ (\ref{dim5}) is
substantially changed under this redefinition that significantly modifies
fermion bilinear terms \ \ \ \ 
\begin{equation}
L_{\overline{\psi }\psi }=i\overline{\psi }\gamma _{\mu }\partial ^{\mu
}\psi +\frac{1}{\mathcal{M}}\partial _{\mu }\overline{\psi }\cdot \partial
^{\mu }\psi -i\Delta e\frac{M}{\mathcal{M}}n_{\mu }\overline{\psi }%
\overleftrightarrow{\partial ^{\mu }}\psi -m_{f}\overline{\psi }\psi 
\label{nl}
\end{equation}%
where we retained the notation $\psi $ for the redefined fermion field and
denoted, as usually, $\overline{\psi }\overleftrightarrow{\partial ^{\mu }}%
\psi =\overline{\psi }(\partial ^{\mu }\psi )-(\partial ^{\mu }\overline{%
\psi })\psi $. Note that the extra fermion derivative terms given in (\ref%
{nl}) is produced just due to the gauge invariance breaking that is
determined by the electromagnetic charge difference $\Delta e=\check{e}-e$
in the total Lagrangian $L_{QED}+$ $L_{\dim 5}$. As a result, there appears
the entirely new, SLIV\ inspired, dispersion relation for a charged fermion
(taken with 4-momentum $p_{\mu }$) of the type%
\begin{equation}
p_{\mu }^{2}\cong \lbrack m_{f}+2\delta (p_{\mu }n^{\mu }/n^{2})]^{2},\text{
\ \ }m_{f}=\left( m-G\frac{M^{2}}{\mathcal{M}}\right) -\delta ^{2}n^{2}%
\mathcal{M}  \label{m}
\end{equation}%
given to an accuracy of $O(m_{f}^{2}/\mathcal{M}^{2})$ with a properly
modified total fermion mass $m_{f}$. Here $\delta $ stands for the small
characteristic, positive or negative, parameter $\delta =(\Delta e)M/%
\mathcal{M}$ of physical Lorentz violation that reflects the joint effect as
is given, from the one hand, by the SLIV scale $M$ and, from the other, by
the charge difference $\Delta e$ being a measure of an internal gauge
non-invariance. Notably, the space-time in itself still possesses Lorentz
invariance, however, fermions with SLIV contributing into their total mass $%
m_{f}$ (\ref{m}) propagate and interact in it in the Lorentz non-covariant
way. At the same time, the photon dispersion relation\ is still retained in
the order $1/\mathcal{M}$ considered.

So, we have shown in the above that SLIV caused by the vector field VEV (\ref%
{vev1}), while\ being\ superficial in gauge invariant theory, becomes
physically significant for some high value of the SLIV\ scale $M$ being
close to the scale $\mathcal{M}$, which is proposed to be located near the
Planck scale $M_{P}$.\ This may happen even at relatively low energies
provided the gauge noninvariance caused by high-dimension couplings of
matter and vector fields is not vanishingly small. This leads, as was
demonstrated in \cite{par}, \ through special dispersion relations appearing
for matter charged fermions, to a new class of phenomena which could be of
distinctive observational interest in particle physics and astrophysics.
They include a significant change in the GZK cutoff for UHE cosmic-ray
nucleons, stability of high-energy pions and $W$ bosons, modification of
nucleon beta decays, and some others just in the presently accessible energy
area in cosmic ray physics.

\ However, though one could speculate about some generically broken or
partial gauge symmetry in a QFT framework \cite{par}, this seems to be too
high price for an actual Lorentz violation which \ may stem from SLIV. And,
what is more, is there really any strong theoretical reason left for Lorentz
invariance to be physically broken, if emergent gauge fields are anyway
generated through the \textquotedblleft safe\textquotedblright\ inactive
SLIV models which recover a conventional Lorentz invariance?

\subsection{Direct Lorentz noninvariant extensions of SM and gravity}

Nevertheless, it must not be ruled out that physical Lorentz invariance
might be explicitly, rather than spontaneously, broken at high energies.
This has attracted considerable attention in recent years as an interesting
phenomenological possibility appearing in direct Lorentz noninvariant
extensions of SM \cite{refs,2,3}. They are generically regarded as being
originated in a more fundamental theory at some large scale probably related
to the Planck scale $M_{P}$. These extensions are in a certain measure
motivated \cite{alan} by a string theory according to which an explicit
(from a QFT point of view) Lorentz violation might be in essence a
spontaneous Lorentz violation related to hypothetical tensor-valued fields
acquiring non-zero VEVs in some non-perturbative vacuum. These VEVs appear
effectively as a set of external background constants so that interactions
with these coefficients have preferred spacetime directions in an effective
QFT framework. The full SM extension (SME) \cite{2} is then defined as the
effective gauge invariant field theory obtained when all such Lorentz
violating vector and tensor field backgrounds are contracted term by term
with SM (and gravitational) fields. However, without a completely viable
string theory, it is not possible to assign definite numerical values to
these coefficients. Moreover, not to have disastrous consequences
(especially when these coefficients are contracted with non-conserved
currents) one also has to additionally propose that observable violating
effects in a low-energy theory with a laboratory scale $m$ should be
suppressed by some power of the ratio $m/M_{P}$ being depended on dimension
of Lorentz breaking couplings. Therefore, one has in this sense a pure
phenomenological approach treating the above arbitrary coefficients as
quantities to be bounded in experiments as if they would simply appear due
to explicit Lorentz violation. Actually, in sharp contrast to the above
formulated SLIV\ in a pure QFT framework, there is nothing in the SME itself
that requires that these Lorentz-violation coefficients emerge due to a
process of a spontaneous Lorentz violation. Indeed, neither the
corresponding massless vector (tensor) NG bosons are required to be
generated, nor these bosons have to be associated with photons or any other
gauge fields of SM.

Apart from Lorentz violation in Standard Model, one can generally think that
the vacuum in quantum gravity may also determine a preferred rest frame at
the microscopic level. If such a frame exists, it must be very much hidden
in low-energy physics since, as was mentioned above, numerous observations
severely limit the possibility of Lorentz violating effects for the SM
fields \cite{refs,2,3}. However, the constraints on Lorentz violation in the
gravitational sector are generally far weaker. This allows to introduce a
pure gravitational Lorentz violation having no significant impact on the SM
physics. An elegant way being close in spirit to our SLIV model (\ref{lag}, %
\ref{gol}) seems to appear in the so called Einstein-aether theory \cite{ted}%
. This is in essence a general covariant theory in which local Lorentz
invariance is broken by some vector \textquotedblleft
aether\textquotedblright\ field $u_{\mu }$ defining the preferred frame.
This field is similar to our constrained vector field $A_{\mu }$, apart from
that this field is taken to be unit $u_{\mu }u^{\mu }=1$. It spontaneously
breaks Lorentz symmetry down to a rotation subgroup, just like as our
constrained vector field $A_{\mu }$ does it for a timelike Lorentz
violation. So, they both give nonlinear realization of Lorentz symmetry thus
leading to its spontaneous violation and induce the corresponding
Goldstone-like modes. The crucial difference is that, while modes related to
the vector field $A_{\mu }$ are collected into the physical photon, modes
associated with the unit vector field $u_{\mu }$ (one helicity-0 and two
helicity-1 modes) exist by them own appearing in some effective SM and
gravitational couplings. Some of them might disappear being absorbed by the
corresponding spin-connection fields related to local Lorentz symmetry in
the Einstein-aether theory. In any case, while aether field $u_{\mu }$ can
significantly change dispersion relations of fields involved, thus leading
to many gravitational and cosmological consequences of preferred frame
effects, it certainly can not be a physical gauge field candidate (say, the
photon in QED).

\subsection{Lorentz violation and supersymmetry. The present paper}

There have been a few active attempts \cite{pos,pos1} over the last decade
to construct Lorentz violating operators for matter and gauge fields in the
supersymmetric Standard Model through their interactions with external
vector and tensor field backgrounds. These backgrounds, according to the SME
approach \cite{2} discussed above, are generated by some Lorentz violating
dynamics at an ultraviolet scale of order the Planck scale. As some
advantages over the ordinary SME, it was shown that in the supersymmetric
Standard Model the lowest possible dimension for such operators is five,
just as we had above in the high-dimensional SLIV case (\ref{dim5}).
Therefore, they are suppressed by at least one power of an ultraviolet
energy scale, providing a possible explanation for the smallness of Lorentz
violation and its stability against radiative corrections. There were
classified all possible dimension five and six Lorentz violating operators
in the SUSY QED \cite{pos1}, analyzed their properties at the quantum level
and described their observational consequences in this theory. These
operators, as was confirmed, do not induce destabilizing $D$-terms, gauge
anomaly and the Chern-Simons term for photons. Dimension-five Lorentz
violating operators were shown to be constrained by low-energy precision
measurements at $10^{-10}-10^{-5}$ level in units of the inverse Planck
scale, while the Planck-scale suppressed dimension six operators are allowed
by observational data.

Also, it has been constructed the supersymmetric extension of the
Einstein-aether theory \cite{puj} discussed above. It has been found that
the dynamics of the super-aether is somewhat richer than of its non-SUSY
counterpart. In particular, the model possesses a family of inequivalent
vacua exhibiting different symmetry breaking patterns while remaining stable
and ghost free. Interestingly enough, as long as the aether VEV preserves
spatial supersymmetry (SUSY algebra without boosts), the Lorentz breaking
does not propagate into the SM sector at the renormalizable level. The
eventual breaking of SUSY, that must be incorporated in any realistic model,
is unrelated to the dynamics of the aether. It is assumed to come from a
different source characterized by a lower energy scale. However, in spite of
its own merits an important final step which would lead to natural
accommodation of this super-aether model into the supergravity framework has
not yet been done.

In contrast, we are strictly focused here on a spontaneous Lorentz violation
in an actual gauge QFT framework related to the Standard Model rather than
in an effective low energy theory with some hypothetical remnants in terms
of external tensor-valued backgrounds originatating somewhere around the
Planck scale. In essence, we try to extend emergent gauge theories with SLIV
and an associated emergence of gauge bosons as massless vector
Nambu-Goldstone modes studied earlier \cite{cfn,jb,kraus,jen,bluhm} (see
also \cite{az,kep,jej,urr,gra}) to their supersymmetric analogs. Generally
speaking, it may turn out that SLIV is not the only reason why massless
photons could dynamically appear, if spacetime symmetry is further enlarged.
In this connection, special interest may be related to supersymmetry, as was
recently argued in \cite{j}. Actually, the situation is changed remarkably
in the SUSY inspired emergent models which, in contrast to non-SUSY
theories, could naturally have some clear observational evidence. Indeed, as
we discussed above (subsection 1.2), ordinary emergent theories admit some
experimental verification only if gauge invariance is properly broken being
caused by some high-dimension couplings. Their SUSY counterparts, and
primarily emergent SUSY QED, are generically appear with supersymmetry being
spontaneously broken in a visible sector to ensure stability of the theory.
Therefore, the verification is now related to an inevitable emergence of a
goldstino-like photino state in the SUSY particle spectrum at low energies,
while physical Lorentz invariant is still left intact\footnote{%
Of course, physical Lorentz violation will also appear if one admits some
gauge noninvariance in the emergent SUSY theory as well. This may happen,
for example, through high-dimension couplings being supersymmetric analogs
of the couplings (\ref{dim5}).}. In this sense, a generic source for
massless photon to appear may be spontaneously broken supersymmetry rather
than physically manifested spontaneous Lorentz violation.

To see how such a scenario may work, we consider supersymmetric QED model
extended by an arbitrary polynomial potential of massive vector superfield
that induces the spontaneous SUSY violation in the visible sector. As a
consequence, a massless photino emerges as the fermion NG mode in the broken
SUSY phase, and a photon as a photino companion to also appear massless in
the tree approximation (section 2). However, the requirement of vacuum
stability in such class of models makes Lorentz invariance to become
spontaneously broken as well. As a consequence, massless photon has now
appeared as the vector NG mode, and also a special gauge invariance is
simultaneously generated in an emergent SUSY QED. This invariance is only
restricted by the supplemented vector field constraint being invariant under
supergauge transformations (section 3). Due to this invariance all
observable SLIV effects appear to be completely cancelled out among
themselves and physical Lorentz invariance is restored. Meanwhile, photino
being mixed with another goldstino appearing from a spontaneous SUSY
violation in the hidden sector largely turns into the light pseudo-goldstino
whose physics seems to be of special observational interest (section 4). And
finally, we conclude (section 5).

\section{Extended supersymmetric QED}

We start by considering a conventional SUSY QED extended by an arbitrary
polynomial potential of \ a general vector superfield $V(x,\theta ,\overline{%
\theta })$ which in the standard parametrization \cite{wess} has a form 
\begin{eqnarray}
V(x,\theta ,\overline{\theta }) &=&C(x)+i\theta \chi -i\overline{\theta }%
\overline{\chi }+\frac{i}{2}\theta \theta S-\frac{i}{2}\overline{\theta }%
\overline{\theta }S^{\ast }  \notag \\
&&-\theta \sigma ^{\mu }\overline{\theta }A_{\mu }+i\theta \theta \overline{%
\theta }\overline{\lambda ^{\prime }}-i\overline{\theta }\overline{\theta }%
\theta \lambda ^{\prime }+\frac{1}{2}\theta \theta \overline{\theta }%
\overline{\theta }D^{\prime },  \label{par}
\end{eqnarray}%
where its vector field component $A_{\mu }$ is usually associated with a
photon. Note that, apart from the conventional photino field $\lambda $ and
the auxiliary $D$ field , the superfield (\ref{par}) contains in general the
additional degrees of freedom in terms of the dynamical $C$ and $\chi $
fields and nondynamical complex scalar field $S$ (we have used the brief
notations, $\lambda ^{\prime }=\lambda +\frac{i}{2}\sigma ^{\mu }\partial
_{\mu }\overline{\chi }$ \ and $D^{\prime }=D+\frac{1}{2}\square C$ with $%
\sigma ^{\mu }=(1,\overrightarrow{\sigma })$ and $\overline{\sigma }^{\mu
}=(1,-\overrightarrow{\sigma })$). The corresponding SUSY invariant
Lagrangian may be written as%
\begin{equation}
\mathcal{L}=L_{SQED}+\sum_{n=1}b_{n}V^{n}|_{D}  \label{slag}
\end{equation}%
where terms in this sum ($b_{n}$ are some constants) for the vector
superfield (\ref{par}) are given through the polynomial $D$-term $V^{n}|_{D}$
expansion into the component fields . It can readily be checked that the
first term in this expansion appears to be the known Fayet-Iliopoulos $D$%
-term, while other terms only contain bilinear, trilinear and quadrilinear
combination of the superfield components $A_{\mu }$, $S$, $\lambda $ and $%
\chi $, respectively\footnote{%
Note that all terms in the sum in (\ref{slag}) except Fayet-Iliopoulos $D$%
-term\ explicitly break gauge invariance which is then recovered in the SUSY
broken phase (see below). For simplicity, we could restrict ourselves to the
third degree superfield polynomial\ potential in the Lagrangian $\mathcal{L}$
(\ref{slag}) to eventually have a theory with dimesionless coupling
constants in interactions of the component fields. However, for completeness
sake, we will proceed with a general superfield potential.}. Actually, there
appear higher-degree terms only for the scalar field component $C(x)$.
Expressing them all in terms of the $C$ field polynomial%
\begin{equation}
P(C)=\sum_{n=1}\frac{n}{2}b_{n}C^{n-1}(x)  \label{pot}
\end{equation}%
and its first three derivatives with respect to the $C$ field 
\begin{equation}
P^{\prime }\equiv \frac{\partial P}{\partial C}\text{ , \ \ }P^{\prime
\prime }\equiv \frac{\partial ^{2}P}{\partial C^{2}}\text{ , \ \ }P^{\prime
\prime \prime }\equiv \frac{\partial ^{3}P}{\partial C^{3}}\text{ }
\label{dd}
\end{equation}%
one has for the whole Lagrangian $\mathcal{L}$ 
\begin{eqnarray}
\mathcal{L} &=&-\text{ }\frac{1}{4}F^{\mu \nu }F_{\mu \nu }+i\lambda \sigma
^{\mu }\partial _{\mu }\overline{\lambda }+\frac{1}{2}D^{2}  \notag \\
&&+\text{ }P\left( D+\frac{1}{2}\square C\right) +P^{\prime }\left( \frac{1}{%
2}SS^{\ast }-\chi \lambda ^{\prime }-\overline{\chi }\overline{\lambda
^{\prime }}-\frac{1}{2}A_{\mu }A^{\mu }\right)  \notag \\
&&+\text{ }\frac{1}{2}P^{\prime \prime }\left( \frac{i}{2}\overline{\chi }%
\overline{\chi }S-\frac{i}{2}\chi \chi S^{\ast }-\chi \sigma ^{\mu }%
\overline{\chi }A_{\mu }\right) +\frac{1}{8}P^{\prime \prime \prime }(\chi
\chi \overline{\chi }\overline{\chi })\text{ .}  \label{lag3}
\end{eqnarray}%
where, for more clarity, we still omitted matter superfields in the model
reserving them for section 4. One can see that the superfield component
fields $C$ and $\chi $ become dynamical due to the potential terms in (\ref%
{lag3}) rather than from the properly constructed supersymmetric field
strengths, as appear for the vector field $A_{\mu }$ and its gaugino
companion $\lambda $. A very remarkable point is that the vector field $%
A_{\mu }$ may only appear with bilinear mass terms in the polynomially
extended Lagrangian (\ref{lag3}). Hence it follows that the
\textquotedblleft bumblebee\textquotedblright\ type model mentioned above (%
\ref{qed1}) with nontrivial vector field potential containing both a
bilinear mass term and a quadrilinear stabilizing term can in no way be
realized in a SUSY context. Meanwhile, the nonlinear QED model, as will
become clear below, successfully matches supersymmetry.

Varying the Lagrangian $\mathcal{L}$ with respect to the $D$ field we come
to 
\begin{equation}
D=-P(C)  \label{d}
\end{equation}%
that finally gives the standard potential energy for for the field system
considered 
\begin{equation}
U(C)=\frac{1}{2}P^{2}\text{ }  \label{pot1}
\end{equation}%
provided that other superfield field components do not develop VEVs. The
potential (\ref{pot1}) may lead to the spontaneous SUSY breaking in the
visible sector if the polynomial $P$ (\ref{pot}) has no real roots, while
its first derivative has, 
\begin{equation}
P\neq 0\text{ , \ }P^{\prime }=0.\text{\ }  \label{der}
\end{equation}%
This requires $P(C)$ to be an even degree polynomial with properly chosen
coefficients $b_{n}$ in (\ref{pot}) that will force its derivative $%
P^{\prime }$ to have at least one root, $C=C_{0}$, in which the potential (%
\ref{pot1}) is minimized and supersymmetry is spontaneously broken. As an
immediate consequence, one can readily see from the Lagrangian $\mathcal{L}$
(\ref{lag3}) that a massless photino $\lambda $ being Goldstone fermion in
the broken SUSY phase make all the other component fields in the superfield $%
V(x,\theta ,\overline{\theta })$ including the photon to also become
massless. However, the question then arises whether this masslessness of
photon will be stable against radiative corrections since gauge invariance
is explicitly broken in the Lagrangian (\ref{lag3}). We show below that it
could be the case if the vector superfield $V(x,\theta ,\overline{\theta })$
would appear to be properly constrained.

\section{Constrained vector superfield}

\subsection{Instability of superfield polynomial potential}

Let us first analyze possible vacuum configurations for the superfield
components in the polynomially extended QED case taken above. In general,
besides the "standard" potential energy expression (\ref{pot1}) determined
solely by the scalar field component $C(x)$ of the vector superfield (\ref%
{par}), one also has to consider other field component contributions into
the potential energy. A possible extension of the potential energy (\ref%
{pot1}) seems to appear only due to the pure bosonic field contributions,
namely due to couplings of the vector and auxiliary scalar fields, $A_{\mu }$
and $S$, in (\ref{lag3}) 
\begin{equation}
\mathcal{U}=\frac{1}{2}P^{2}+\frac{1}{2}P^{\prime }(A_{\mu }A^{\mu
}-SS^{\ast })\text{ }  \label{pot1a}
\end{equation}%
rather than due to the potential terms containing the superfield fermionic
components\footnote{%
Actually, this restriction is not essential for what follows and is taken
just for simplicity. Generally, the fermion bilinears involved could also
develop VEVs.}. It can be immediately seen that these new couplings in (\ref%
{pot1a}) can make the potential unstable since the vector and scalar fields
mentioned may in general develop any arbitrary VEVs. This happens, as
emphasized above, due the fact that their bilinear term contributions are
not properly compensated by appropriate four-linear field terms which are
generically absent in a SUSY theory context.

For more detail we consider the extremum conditions for the entire potential
(\ref{pot1a}) with respect to all fields involved: $C$, $A_{\mu }$ and $S$.
They are given by the appropriate first partial derivative equations%
\begin{eqnarray}
\mathcal{U}_{C}^{\prime } &=&PP^{\prime }+\frac{1}{2}P^{\prime \prime
}(A_{\mu }A^{\mu }-SS^{\ast })=0,\text{ }  \notag \\
\mathcal{U}_{A_{\mu }}^{\prime } &=&P^{\prime }A^{\mu }=0,\text{ \ \ }%
\mathcal{U}_{S}^{\prime }=-P^{\prime }S^{\ast }=0.  \label{"}
\end{eqnarray}%
where and hereafter all the VEVs are denoted by the corresponding field
symbols (supplied below with the lower index $0$). One can see that there
can occur a local minimum for the potential (\ref{pot1a}) with the unbroken
SUSY solution\footnote{%
Hereafter by $P(C_{0})$ and $P^{\prime }(C_{0})$ are meant the $C$ field
polynomial $P$ (\ref{pot}) and its functional derivative $P^{\prime }$ (\ref%
{dd}) taken in the potential extremum point $C_{0}$.} 
\begin{equation}
C=C_{0},\text{ }P(C_{0})=0\text{, }P^{\prime }(C_{0})\neq 0\text{ };\text{ \ 
}A_{\mu 0}=0,\text{ \ }S_{0}=0  \label{sol}
\end{equation}%
with the vanishing potential energy 
\begin{equation}
\mathcal{U}_{\min }^{s}=0  \label{pe}
\end{equation}%
provided that the polynomial $P$ (\ref{pot}) has some real root $C=C_{0}$.
Otherwise, a local minimum with the broken SUSY solution can occur for some
other $C$ field value (though denoted by the same letter $C_{0}$) 
\begin{equation}
C=C_{0},\text{ }P(C_{0})\neq 0\text{, }P^{\prime }(C_{0})=0\text{ };\text{ \ 
}A_{\mu 0}\neq 0,\text{ \ }S_{0}\neq 0,\text{ }A_{\mu 0}A_{0}^{\mu
}-S_{0}S_{0}^{\ast }=0  \label{sol2}
\end{equation}%
In this case one has the non-zero potential energy 
\begin{equation}
\mathcal{U}_{\min }^{as}=\frac{1}{2}[P(C_{0})]^{2}  \label{pe2}
\end{equation}%
as directly follows from the extremum equations (\ref{"}) and potential
energy expression (\ref{pot1a}).

However, as shows the standard second partial derivative test, the fact is
that the local minima mentioned above are minima with respect to the $C$
field VEV ($C_{0}$) only. Actually, for all three fields VEVs included the
potential (\ref{pot1a}) has indeed saddle points with "coordinates"
indicated in (\ref{sol}) and (\ref{sol2}), respectively. For a testing
convenience this potential can be rewritten in the form 
\begin{equation}
\mathcal{U}=\frac{1}{2}P^{2}+\frac{1}{2}P^{\prime }g^{\Theta \Theta ^{\prime
}}B_{\Theta }B_{\Theta ^{\prime }}\text{ , \ }g^{\Theta \Theta ^{\prime
}}=diag\text{ }(1,-1,-1,-1,-1,-1)\text{\ \ \ }  \label{pot1b}
\end{equation}%
with only two variable fields $C$ and $B_{\Theta }$ where the new field $%
B_{\Theta }$ unifies the $A_{\mu }$ and $S$ field components, $B_{\Theta
}=(A_{\mu },S_{a})$ ($\Theta =\mu ,a;$ $\mu =0,1,2,3$; $a=1,2$)\footnote{%
Interestingly, the $B_{\Theta }$\ term in the potential (\ref{pot1b})
possesses the accidental $SO(1,5)$ symmetry. This symmetry, though it is not
shared by kinetic terms, appears in fact to be stable under radiative
corrections since $S$ field is non-dynamical and, therefore, can always be
properly arranged.}. The complex $S$ field is now taken in a real basis, $%
S_{1}=(S+S^{\ast })/\sqrt{2}$ and $S_{2}=(S-S^{\ast })/i\sqrt{2}$, so that
the "vector" $B_{\Theta }$ field has one time and five space components. As
a result, one finally comes to the following Hessian $7\times 7$ matrix
(being in fact the second-order partial derivatives matrix taken in the
extremum point ($C_{0}$, $A_{\mu 0}$, $S_{0}$) (\ref{sol})) 
\begin{equation}
H(\mathcal{U}^{s})=\left[ 
\begin{array}{cc}
\lbrack P^{\prime }(C_{0})]^{2} & 0 \\ 
0 & P^{\prime }(C_{0})g^{\Theta \Theta ^{\prime }}%
\end{array}%
\right] \text{, \ }\left\vert H(\mathcal{U}^{s})\right\vert =-\text{\ }%
[P^{\prime }(C_{0})]^{8}\text{ .}  \label{hess}
\end{equation}%
This matrix clearly has the negative determinant $\left\vert H(\mathcal{U}%
^{s})\right\vert $, as is indicated above, that confirms that the potential
definitely has a saddle point for the solution (\ref{sol}). This means the
VEVs of the $A_{\mu }$ and $S$ fields can take in fact any arbitrary value
making the potential (\ref{pot1a}, \ref{pot1b}) to be unbounded from below
in the unbroken SUSY case that is certainly inaccessible.

One might think that in the broken SUSY case the situation would be better
since due to the conditions (\ref{sol2}) the $B_{\Theta }$ term completely
disappears from the potential $\mathcal{U}$ (\ref{pot1a}, \ref{pot1b}) in
the ground state. Unfortunately, the direct second partial derivative test
in this case is inconclusive since the determinant of the corresponding
Hessian $7\times 7$ matrix appears to vanish 
\begin{equation}
H(\mathcal{U}^{as})=\left[ 
\begin{array}{cc}
P(C_{0})P^{\prime \prime }(C_{0}) & P^{\prime \prime }(C_{0})g^{\Theta
\Theta ^{\prime }}B_{\Theta ^{\prime }} \\ 
P^{\prime \prime }(C_{0})g^{\Theta \Theta ^{\prime }}B_{\Theta ^{\prime }} & 
0%
\end{array}%
\right] \text{, \ \ }\left\vert H(\mathcal{U}^{as})\right\vert =0\text{ .}
\label{hess2}
\end{equation}%
Nevertheless, since in general the $B_{\Theta }$ term can take both positive
and negative values in small neighborhoods around the vacuum point ($C_{0}$, 
$A_{\mu 0}$, $S_{0}$) where the conditions (\ref{sol2}) are satisfied, this
point is also turned out to be a saddle point. Thus, the potential $\mathcal{%
U}$ (\ref{pot1a}, \ref{pot1b}) appears generically unstable both in SUSY
invariant and SUSY broken phase.

\subsection{Stabilization of vacuum by constraining vector superfield}

The only possible way to stabilize the ground states (\ref{sol}) and (\ref%
{sol2}) seems to seek the proper constraints on the superfield component
fields ($C$, $A_{\mu }$, $S$) themselves rather than on their expectation
values. Indeed, if such (potential bounding) constraints are physically
realizable, the vacua (\ref{sol}) and (\ref{sol2}) will be automatically
stabilized.

In a SUSY context a constraint can only be put on the entire vector
superfield $V(x,\theta ,\overline{\theta })$ (\ref{par}) rather than
individually on its field components. Actually, we can constrain our vector
superfield $V(x,\theta ,\overline{\theta })$ by analogy with the constrained
vector field in the nonlinear QED model (see (\ref{lag})). This will be done
again through some invariant Lagrange multiplier coupling simply adding its $%
D$ term to the above Lagrangian (\ref{slag}, \ref{lag3}) 
\begin{equation}
\mathcal{L}_{tot}=\mathcal{L}+\frac{1}{2}{\large \Lambda }(V-C_{0})^{2}|_{D}%
\text{ ,}  \label{ext}
\end{equation}%
where ${\large \Lambda }(x,\theta ,\overline{\theta })$ is some auxiliary
vector superfield, while $C_{0}$ is the constant background value of the $C$
field for which potential $U$ (\ref{pot1}) vanishes as is required for the
supersymmetric minimum or has some nonzero value corresponding to the SUSY
breaking minimum (\ref{der}) in the visible sector. We will consider both
cases simultaneously using the same notation $C_{0}$ for either of the
potential minimizing values of the $C$ field.

Note first of all, the Lagrange multiplier term in (\ref{ext}) has in fact
the simplest possible form that leads to some nontrivial constrained
superfield $V(x,\theta ,\overline{\theta })$. The alternative minimal forms,
such as the bilinear form ${\large \Lambda }(V-C_{0})$ or trilinear one $%
{\large \Lambda }(V^{2}-C_{0}^{2})$, appear too restrictive. One can easily
confirm that they eliminate most component fields in the superfield $%
V(x,\theta ,\overline{\theta })$ including the physical photon and photino
fields that is definitely inadmissible. As to appropriate non-minimal high
linear multiplier forms, they basically lead to the same consequences as
follow from the minimal multiplier term taken in the total Lagrangian (\ref%
{ext}). Writing down its invariant $D$ term through the component fields one
finds 
\begin{eqnarray}
{\large \Lambda }(V-C_{0})^{2}|_{D} &=&{\large C}_{{\large \Lambda }}\left[ 
\widetilde{C}D^{\prime }+\left( \frac{1}{2}SS^{\ast }-\chi \lambda ^{\prime
}-\overline{\chi }\overline{\lambda ^{\prime }}-\frac{1}{2}A_{\mu }A^{\mu
}\right) \right]  \notag \\
&&+\text{ }{\large \chi }_{{\large \Lambda }}\left[ 2\widetilde{C}\lambda
^{\prime }+i(\chi S^{\ast }+i\sigma ^{\mu }\overline{\chi }A_{\mu })\right] +%
\overline{{\large \chi }}_{{\large \Lambda }}[2\widetilde{C}\overline{%
\lambda ^{\prime }}-i(\overline{\chi }S-i\chi \sigma ^{\mu }A_{\mu })] 
\notag \\
&&+\text{ }\frac{1}{2}{\large S}_{{\large \Lambda }}\left( \widetilde{C}%
S^{\ast }+\frac{i}{2}\overline{\chi }\overline{\chi }\right) +\frac{1}{2}%
{\large S}_{{\large \Lambda }}^{\ast }\left( \widetilde{C}S-\frac{i}{2}\chi
\chi \right)  \notag \\
&&+\text{ }2{\large A}_{{\large \Lambda }}^{\mu }(\widetilde{C}A_{\mu }-\chi
\sigma _{\mu }\overline{\chi })+2{\large \lambda }_{{\large \Lambda }%
}^{\prime }(\widetilde{C}\chi )+2\overline{{\large \lambda }}_{{\large %
\Lambda }}^{\prime }(\widetilde{C}\overline{\chi })+\frac{1}{2}{\large D}_{%
{\large \Lambda }}^{\prime }\widetilde{C}^{2}  \label{lm1}
\end{eqnarray}%
where 
\begin{equation}
{\large C}_{{\large \Lambda }},\text{ }{\large \chi }_{{\large \Lambda }},%
\text{ }{\large S}_{{\large \Lambda }},\text{ }{\large A}_{{\large \Lambda }%
}^{\mu },\text{ }{\large \lambda }_{{\large \Lambda }}^{\prime }={\large %
\lambda }_{{\large \Lambda }}+\frac{i}{2}\sigma ^{\mu }\partial _{\mu }%
\overline{{\large \chi }}_{{\large \Lambda }},\text{ }{\large D}_{{\large %
\Lambda }}^{\prime }={\large D}_{{\large \Lambda }}+\frac{1}{2}\square 
{\large C}_{{\large \Lambda }}  \label{comp}
\end{equation}%
are the component fields of the Lagrange multiplier superfield ${\large %
\Lambda }(x,\theta ,\overline{\theta })$ in the standard parametrization (%
\ref{par}) and $\widetilde{C}$ stands for the difference $C(x)-C_{0}$.
Varying the Lagrangian (\ref{ext}) with respect to these fields and properly
combining their equations of motion 
\begin{equation}
\frac{\partial \mathcal{L}_{tot}}{\partial \left( {\large C}_{{\large %
\Lambda }},{\large \chi }_{{\large \Lambda }},{\large S}_{{\large \Lambda }},%
{\large A}_{{\large \Lambda }}^{\mu },{\large \lambda }_{{\large \Lambda }},%
{\large D}_{{\large \Lambda }}\right) }=0
\end{equation}%
we find the constraints which appear to put on the $V$ superfield components 
\begin{equation}
C=C_{0},\text{ \ }\chi =0,\text{\ \ }A_{\mu }A^{\mu }=SS^{\ast }\text{.}
\label{const1}
\end{equation}%
Again, as before in non-SUSY case (\ref{lag}), we only take a solution with
initial values for all fields (and their momenta) chosen so as to restrict
the phase space to vanishing values of the multiplier component fields (\ref%
{comp}) that will provide a ghost-free theory with a positive Hamiltonian%
\footnote{%
As in the non-supersymmetric case discussed above (see footnote$^{1}$), this
solution with all vanishing components of the basic Lagrangian multiplier
superfield ${\large \Lambda }(x,\theta ,\overline{\theta })$ can be reached
by introducing some extra Lagrange multiplier term.}.

Remarkably, the constraints (\ref{const1}) does not touch the physical
degrees of freedom of the \ superfield $V(x,\theta ,\overline{\theta })$
related to photon and photino fields. The point is, however, that\qquad
apart from the constraints (\ref{const1}), one has the equations of motion
for all fields involved in the basic superfield $V(x,\theta ,\overline{%
\theta })$. With vanishing multiplier component fields (\ref{comp}), as was
proposed above, these equations appear in fact as extra constraints on
components of the superfield $V(x,\theta ,\overline{\theta })$. Indeed,
equations of motion for the fields $C$, $S$ and ${\large \chi }$ received by
the corresponding variations of the total Lagrangian $\mathcal{L}$ (\ref%
{lag3}) are turned out to be, respectively, 
\begin{equation}
P(C_{0})P^{\prime }(C_{0})=0,\text{ \ }S(x)P^{\prime }(C_{0})=0\text{ , \ }%
\lambda (x)P^{\prime }(C_{0})=0  \label{nc}
\end{equation}%
where the basic constraints (\ref{const1}) emerging at the potential
extremum point $C=C_{0}$ have also been used. One can immediately see now
that these equations turn to trivial identities in the broken SUSY case, in
which the factor $P^{\prime }(C_{0})$ in each of them appears to be
identically vanished, $P^{\prime }(C_{0})$ $=0$ (\ref{sol2}). In the
unbroken SUSY case, in which the potential (\ref{pot1}) vanishes instead,
i.e. $P(C_{0})=0$ (\ref{sol}), the situation is drastically changed. Indeed,
though the first equation in (\ref{nc}) still automatically turns into
identity at the extremum point $C(x)=C_{0}$, other two equations require
that the auxiliary field $S$ and photino field $\lambda $ have to be
identically vanished as well. This causes in turn that the photon field
should also be vanished according to the basic constraints (\ref{const1}).
Besides, the $D$ field component in the vector superfield is also vanished
in the unbroken SUSY case according to the equation (\ref{d}), $%
D=-P(C_{0})=0 $. Thus, one is ultimately left with a trivial superfield $%
V(x,\theta ,\overline{\theta })$ which only contains the constant $C$ field
component $C_{0}$ that is unacceptable. So, we have to conclude that the
unbroken SUSY fails to provide stability of the potential (\ref{pot1a}) even
by constraining the superfield $V(x,\theta ,\overline{\theta })$. In
contrast, in the spontaneously broken SUSY case extra constraints do not
appear at all, and one has a physically meaningful theory that we basically
consider in what follows.

Actually, substituting the constraints (\ref{const1}) into the total
Lagrangian $\mathcal{L}_{tot}$ (\ref{ext}, \ref{lag3}) we eventually come to
the emergent SUSY QED appearing in the broken SUSY phase%
\begin{equation}
\mathcal{L}_{tot}^{^{\mathbf{em}}}=-\text{ }\frac{1}{4}F^{\mu \nu }F_{\mu
\nu }+i\lambda \sigma ^{\mu }\partial _{\mu }\overline{\lambda }+\frac{1}{2}%
D^{2}+P(C_{0})D\text{ , \ }A_{\mu }A^{\mu }=SS^{\ast }  \label{fin}
\end{equation}%
supplemented by the vector field constraint as its vacuum stability
condition. Remarkably, for the constrained vector superfield involved 
\begin{equation}
\widehat{V}(x,\theta ,\overline{\theta })=C_{0}+\frac{i}{2}\theta \theta S-%
\frac{i}{2}\overline{\theta }\overline{\theta }S^{\ast }-\theta \sigma ^{\mu
}\overline{\theta }A_{\mu }+i\theta \theta \overline{\theta }\overline{%
\lambda }-i\overline{\theta }\overline{\theta }\theta \lambda +\frac{1}{2}%
\theta \theta \overline{\theta }\overline{\theta }D,  \label{sup}
\end{equation}%
we have the almost standard SUSY QED Lagrangian with the same states -
photon, photino and an auxiliary scalar $D$ field - in its gauge
supermultiplet, while another auxiliary complex scalar field $S$ gets only
involved in the vector field constraint. The linear (Fayet-Iliopoulos) $D$%
-term with the effective coupling constant $P(C_{0})$ in (\ref{fin}) shows
that the supersymmetry in the theory is spontaneously broken due to which
the $D$ field acquires the VEV, $D=-P(C_{0})$. Taking the nondynamical $S$
field in the constraint (\ref{const1}) to be some constant background field
(for a more formal discussion, see below) we come to the SLIV\ constraint (%
\ref{const}) which we discussed above regarding an ordinary
non-supersymmetric QED theory (section1). As is seen from this constraint in
(\ref{fin}), one may only have the time-like SLIV in a SUSY framework but
never the space-like one. There also may be a light-like SLIV, if the $S$
field vanishes\footnote{%
Indeed, this case, first mentioned in \cite{nambu}, may also mean
spontaneous Lorentz violation with a nonzero VEV $<A_{\mu }>$ $=(\widetilde{M%
},0,0,\widetilde{M})$ and Goldstone modes $A_{1,2}$ and $(A_{0}+A_{3})/2$\ $-%
\widetilde{M}.$ The "effective" Higgs mode $(A_{0}-A_{3})/2$ can be then
expressed through Goldstone modes so as the light-like condition $A_{\mu
}^{2}=0$ to be satisfied.}. So, any possible choice for the $S$ field
corresponds to the particular gauge choice for the vector field $A_{\mu }$
in an otherwise gauge invariant theory.

\subsection{Constrained superfield: a formal view}

We conclude this section by showing that the extended Lagrangian $\mathcal{L}%
_{tot}$ (\ref{ext}, \ref{lag3}), underlying the emergent QED model described
above, as well as the vacuum stability constraints on the superfield
component fields (\ref{const1}) appearing due to the Lagrange multiplier
term in (\ref{ext}) are consistent with supersymmetry. The first part of
this assertion is somewhat immediate since the Lagrangian $\mathcal{L}_{tot}$%
, aside from the standard supersymmetric QED part $L_{SQED}$ (\ref{slag}),
only contains $D$-terms of various vector superfield products. They are, by
definition, invariant under conventional SUSY transformations \cite{wess}
which for the component fields (\ref{par}) of a general superfield $%
V(x,\theta ,\overline{\theta })$ (\ref{par}) are written as 
\begin{eqnarray}
\delta _{\xi }C &=&i\xi \chi -i\overline{\xi }\overline{\chi }\text{ , \ }%
\delta _{\xi }\chi =\xi S+\sigma ^{\mu }\overline{\xi }(\partial _{\mu
}C+iA_{\mu })\text{ , \ }\frac{1}{2}\delta _{\xi }S=\overline{\xi }\overline{%
\lambda }+\overline{\sigma }_{\mu }\partial ^{\mu }\chi \text{ ,}  \notag \\
\delta _{\xi }A_{\mu } &=&\xi \partial _{\mu }\chi +\overline{\xi }\partial
_{\mu }\overline{\chi }+i\xi \sigma _{\mu }\overline{\lambda }-i\lambda
\sigma _{\mu }\overline{\xi }\text{ , \ }\delta _{\xi }\lambda =\frac{1}{2}%
\xi \sigma ^{\mu }\overline{\sigma }^{\nu }F_{\mu \nu }+\xi D\text{ ,} 
\notag \\
\delta _{\xi }D &=&-\xi \sigma ^{\mu }\partial _{\mu }\overline{\lambda }+%
\overline{\xi }\sigma ^{\mu }\partial _{\mu }\lambda \text{ .}  \label{trans}
\end{eqnarray}%
However, there may still be left a question whether supersymmetry remains in
force when the constraints (\ref{const1})\ on the field space are "switched
on" thus leading to the final Lagrangian $\mathcal{L}_{tot}^{^{\mathbf{em}}}$
(\ref{fin}) in the broken SUSY phase with both dynamical fields $C$ and $%
\chi $ eliminated. This Lagrangian appears similar to the standard
supersymmetric QED taken in the Wess-Zumino gauge, except that the
supersymmetry is spontaneously broken in our case. In both cases the photon
stress tensor $F_{\mu \nu }$, photino $\lambda $ and nondynamical scalar $D$
field form an irreducible representation of the supersymmetry algebra (the
last two line in (\ref{trans})). Nevertheless, any reduction of component
fields in the vector superfield is not consistent in general with the linear
superspace version of supersymmetry transformations, whether it is the
Wess-Zumino gauge case or our constrained superfield $\widehat{V}$ (\ref{sup}%
). Indeed, a general SUSY transformation does not preserve the Wess-Zumino
gauge: a vector superfield in this gauge acquires some extra terms when
being SUSY transformed. The same occurs with our constrained superfield $%
\widehat{V}$ as well. The point, however, is that in both cases a total
supergauge transformation 
\begin{equation}
V\rightarrow V+i(\Omega -\Omega ^{\ast })\text{ ,}  \label{spg}
\end{equation}%
where $\Omega $ is a chiral superfield gauge transformation parameter, can
always restore a superfield initial form. Actually, the only difference
between these two cases is that whereas the Wess-Zumino supergauge leaves an
ordinary gauge freedom untouched, in our case this gauge is unambiguously
fixed in terms of the above \ vector field constraint (\ref{const1}).
However, this constraint remains under supergauge transformation (\ref{spg})
applied to our superfield $\widehat{V}$ (\ref{sup}). Indeed, the essential
part of this transformation which directly acts on the constraint (\ref%
{const1}) has the form 
\begin{equation}
\widehat{V}\rightarrow \widehat{V}+i\theta \theta F-i\overline{\theta }%
\overline{\theta }F^{\ast }-2\theta \sigma ^{\mu }\overline{\theta }\partial
_{\mu }\varphi \text{ .}  \label{sg}
\end{equation}%
where the real and complex scalar field components, $\varphi $ and $F$, in a
chiral superfield parameter $\Omega $ \ are properly activated. As a result,
the vector and scalar fields, $A_{\mu }$ and $S$, in the supermultiplet $%
\widehat{V}$ (\ref{sup}) transform as 
\begin{equation}
A_{\mu }\rightarrow A_{\mu }^{\prime }=A_{\mu }-\partial _{\mu }(2\varphi )%
\text{ , \ \ }S\rightarrow S^{\prime }=S+2F\text{ .}  \label{gra}
\end{equation}%
It can be immediately seen that our basic Lagrangian $\mathcal{L}_{tot}^{^{%
\mathbf{em}}}$ (\ref{fin}) being gauge invariant and containing no the
scalar field $S$ is automatically invariant under either of these two
transformations individually. In contrast, the supplementary vector field
constraint (\ref{const1}), though it is also turned out to be invariant
under supergauge transformations (\ref{gra}), but only if they are made
jointly. Indeed, for any choice of the scalar $\varphi $ in (\ref{gra})
there can always be found such a scalar $F$ (and vice versa) that the
constraint remains invariant%
\begin{equation}
A_{\mu }A^{\mu }=SS^{\ast }\rightarrow A_{\mu }^{\prime }A^{\prime \mu
}=S^{\prime }S^{\prime \ast }  \label{sc1}
\end{equation}%
In other words, the vector field constraint is invariant under supergauge
transformations (\ref{gra}) but not invariant under an ordinary gauge
transformation. As a result, in contrast to the Wess-Zumino case, the
supergauge fixing in our case will also lead to the ordinary gauge fixing.
We will use this supergauge freedom to reduce the $S$ field to some constant
background value and find the final equation for the gauge function $\varphi
(x)$. So, for the parameter field $F$ chosen in such a way to have 
\begin{equation}
S^{\prime }=S+2F=Me^{i\alpha (x)}\text{ },  \label{1}
\end{equation}%
where $M$ is some constant mass parameter (and $\alpha (x)$ is an arbitrary
phase), we come in (\ref{sc1}) to 
\begin{equation}
(A_{\mu }-2\partial _{\mu }\varphi )(A^{\mu }-2\partial ^{\mu }\varphi
)=M^{2}\text{ .}  \label{2}
\end{equation}%
that is precisely our old SLIV constraint (\ref{const}) being varied by the
gauge transformation (\ref{gra}). Recall that this constraint, as was
thoroughly discussed in Introduction (subsection 1.1), only fixes gauge (to
which such a gauge function $\varphi (x)$ has to satisfy), rather than
physically breaks gauge invariance. Notably, in contrast to the \ non-SUSY
case where this constraint was merely postulated, it now follows from the
vacuum stability and supergauge invariance in the emergent SUSY QED.
Besides, this constraint, as mentioned above, may only be time-like (and
light-like if the mass parameter $M$ is taken to be zero). When such
inactive time-like SLIV is properly developed one come to the essentially
nonlinear emergent SUSY QED\ in which the physical photon arises as a
three-dimensional Lorentzian NG mode (just as is in non-SUSY case for the
time-like SLIV, see subsection 1.1).

To finalize, it was shown that the vacuum stability constraints\ (\ref%
{const1}) on the allowed configurations of the physical fields in a general
polynomially extended Lagrangian (\ref{ext}) appear entirely consistent with
supersymmetry. In the broken SUSY phase one eventually comes to the standard
SUSY QED type Lagrangian (\ref{fin}) being supplemented by the vector field
constraint which is invariant under supergauge transformations. One might
think that, unlike the gauge invariant linear (Fayet-Iliopoulos) superfield
term, the quadratic and higher order superfield terms in the starting
Lagrangian (\ref{ext}) would seem to break gauge invariance. However, this
fear proves groundless. Actually, as was shown above, this breaking amounts
to the gauge fixing determined by the nonlinear vector field constraint (\ref%
{sc1}). It is worth noting that this constraint formally follows from the
SUSY invariant Lagrange multiplier term in (\ref{ext}) for which is required
the phase space to be restricted to vanishing values of all the multiplier
component fields (\ref{comp}). The total vanishing of the multiplier
superfield provides the SUSY invariance of such restrictions. Any non-zero
multiplier component field left in the Lagrangian would immediately break
supersymmetry and, even worse, would eventually lead to ghost modes in the
theory and a Hamiltonian unbounded from below.

\section{Broken SUSY phase: photino as pseudo-goldstino}

Let us now turn to matter superfields which have not yet been included in
the model. In their presence spontaneous SUSY breaking in the visible
sector, which fundamentally underlies our approach, might be
phenomenologically ruled out by the well-known supertrace sum rule \cite%
{wess} for actual masses of quarks and leptons and their superpartners%
\footnote{%
Note that an inclusion of direct soft mass terms for scalar superpartners in
the model would mean in general that the visible SUSY sector is explicitly,
rather than spontaneosly, broken that could immediately invalidate the whole
idea of the massless photons as the zero Lorentzian modes triggered by the
spontaneously broken supersymmetry.}. However, this sum rule is acceptably
relaxed when taking into account large radiative corrections to masses of
supersymmetric particles that proposedly stems from the hidden sector. This
is just what one may expect in conventional supersymmetric theories with the
standard two-sector paradigm, according to which SUSY breaking entirely
occurs in a hidden sector and then this spontaneous breaking is mediated to
the visible sector by some indirect interactions whose nature depends on a
particular mediation scenario \cite{wess}. An emergent QED approach
advocated here requires some modification of this idea in such a way that,
while a hidden sector is largely responsible for spontaneous SUSY breaking,
supersymmetry can also be spontaneously broken in the visible sector that
ultimately leads to a double spontaneous SUSY breaking pattern.

We may suppose, just for uniformity, only $D$-term SUSY breaking both in the
visible and hidden sectors\footnote{%
In general, both $D$- and $F$-type terms can be simultaneously used in the
visible and hidden sectors\ (usually just $F$-term SUSY breaking is used in
both sectors \cite{wess}).}. Properly, our supersymmetric QED model may be
further extended by some extra local $U^{\prime }(1)$ symmetry which is
proposed to be broken at very high energy scale $M^{\prime }$ (for some
appropriate anomaly mediated scenario, see \cite{ross} and references
therein). It is natural to think that due to the decoupling theorem all
effects of the $U^{\prime }(1)$ are suppressed at energies \ $E<<M^{\prime }$
by powers of $1/M^{\prime }$ and only the $D^{\prime }$-term of the
corresponding vector superfield $V^{\prime }(x,\theta ,\overline{\theta })$
remains in essence when going down to low energies. Actually, this term with
a proper choice of messenger fields and their couplings naturally provides
the $M_{SUSY\text{ \ }}$order contributions to masses of scalar
superpartners.

As a result, the simplified picture discussed above (in sections 2 and 3) is
properly changed: a strictly massless fermion eigenstate, the true goldstino 
$\zeta _{g}$, should now be some mix of the visible sector photino $\lambda $
and the hidden sector\ goldstino $\lambda ^{\prime }$%
\begin{equation}
\zeta _{g}=\frac{\left\langle D\right\rangle \lambda +\left\langle D^{\prime
}\right\rangle \lambda ^{\prime }}{\sqrt{\left\langle D\right\rangle
^{2}+\left\langle D^{\prime }\right\rangle ^{2}}}\text{ .}  \label{tr}
\end{equation}%
where $\left\langle D\right\rangle $ and $\left\langle D^{\prime
}\right\rangle $ are the corresponding $D$-component VEVs in the visible and
hidden sectors, respectively. Another orthogonal combination of them may be
referred to as the pseudo-goldstino $\zeta _{pg}$, 
\begin{equation}
\zeta _{pg}=\frac{\left\langle D^{\prime }\right\rangle \lambda
-\left\langle D\right\rangle \lambda ^{\prime }}{\sqrt{\left\langle
D\right\rangle ^{2}+\left\langle D^{\prime }\right\rangle ^{2}}}\text{ .}
\label{ps}
\end{equation}%
In the supergravity context, the true goldstino $\zeta _{g}$ is eaten
through the super-Higgs mechanism to form the longitudinal component of the
gravitino, while the pseudo-goldstino $\zeta _{pg}$ gets some mass
proportional to the gravitino mass from supergravity effects. Due to large
soft masses required to be mediated, one may generally expect that SUSY is
much stronger broken in the hidden sector than in the visible one, $%
\left\langle D^{\prime }\right\rangle >>$ $\left\langle D\right\rangle $,
that means in turn the pseudo-goldstino $\zeta _{pg}$ is largely the photino 
$\lambda ,$ 
\begin{equation}
\zeta _{pg}\simeq \lambda \text{ .}
\end{equation}%
These pseudo-goldstonic photinos seem to be of special observational
interest in the model that, apart from some indication of the QED emergence
nature, may shed light on SUSY breaking physics. The possibility that the
supersymmetric Standard Model visible sector might also spontaneously break
SUSY thus giving rise to some pseudo-goldstino state was also considered,
though in a different context, in \cite{vis,tha}.

Interestingly enough, our polynomially extended SQED Lagrangian (\ref{slag})
is not only SUSY invariant but also generically possesses a continuous $R$%
-symmetry $U(1)_{R}$ \cite{wess}. Indeed, vector superfields always have
zero $R$-charge, since they are real. Accordingly, it follows that the
physical field components in the constrained vector superfield $\widehat{V}$
(\ref{sup}) transform as%
\begin{equation}
A_{\mu }\rightarrow A_{\mu }\text{ , \ }\lambda \rightarrow e^{i\alpha
}\lambda \text{ , \ }D\rightarrow D  \label{tran}
\end{equation}%
and so have $R$ charges $0$, $1$ and $0$, respectively. Along with that, we
assume a suitable $R$-symmetric matter superfield setup as well making a
proper $R$-charge assignment for basic fermions and scalars (and messenger
fields) involved. This will lead to the light pseudo-goldstino in the
gauge-mediated scenario. Indeed, if the visible sector possesses an $R$%
-symmetry which is preserved in the course of mediation the pseudo-goldstino
mass is protected up to the supergravity effects which violate an $R$%
-symmetry. As a result, the pseudo-goldstino mass appears proportional to
the gravitino mass, and, eventually, the same region of parameter space
simultaneously solves both gravitino and pseudo-goldstino overproduction
problems in the early universe \cite{tha}.

Apart from cosmological problems, many other sides of new physics related to
pseudo-goldstinos appearing through the multiple SUSY breaking were also
studied recently (see \cite{vis,tha,gol} and references therein). The point,
however, is that there have been exclusively used non-vanishing $F$-terms as
the only mechanism of the visible SUSY breaking in models considered. In
this connection, our pseudo-goldstonic photinos solely caused by
non-vanishing $D$-terms in the visible SUSY sector may lead to somewhat
different observational consequences. One of the most serious differences
may be related to the Higgs boson decays when the present SUSY QED is
further extended to the supersymmetric Standard Model. For the
cosmologically safe masses of pseudo-goldstino and gravitino ($\lesssim $ $%
1keV$, as typically follows from the $R$-symmetric gauge mediation) these
decays are appreciably modified. Actually, the dominant channel becomes the
conversion of the Higgs boson (say, the lighter CP-even Higgs boson $h^{0}$)
into a conjugated pair of corresponding pseudo-sgoldstinos $\phi _{pg}$ and $%
\overline{\phi }_{pg}$ (being superpartners of\ pseudo-goldstinos $\zeta
_{pg}$ and $\overline{\zeta }_{pg}$, respectively), 
\begin{equation}
h^{0}\rightarrow \phi _{pg}+\overline{\phi }_{pg}\text{ ,}
\end{equation}
once it is kinematically allowed. This means that the Higgs boson will
dominantly decay invisibly for $F$-term SUSY breaking in a visible sector 
\cite{tha}. By contrast, for the $D$-term SUSY breaking case considered here
the roles of pseudo-goldstino and pseudo-sgoldstino are just played by
photino and photon, respectively, that could make the standard two-photon
decay channel of the Higgs boson to be even somewhat enhanced. In the light
of recent discovery of the Higgs-like state \cite{h} just through its
visible decay modes, the $F$-term SUSY breaking in the visible sector seems
to be disfavored by data, while $D$-term SUSY breaking is not in trouble
with them. \ 

\section{Concluding remarks}

It is well known that spontaneous Lorentz violation in general vector field
theories may lead to an appearance of massless Nambu-Goldstone modes which
are identified with photons and other gauge fields in the Standard Model.
Nonetheless, it may turn out that SLIV is not the only reason for emergent
massless photons to appear, if spacetime symmetry is further enlarged. In
this connection, a special interest may be related to supersymmetry and its
possible theoretical and observational relation to SLIV.

To see how such a scenario may work we have considered supersymmetric QED
model extended by an arbitrary polynomial potential of a general vector
superfield $V(x,\theta ,\overline{\theta })$ whose pure vector field
component $A_{\mu }(x)$ is associated with a photon in the Lorentz invariant
phase. Gauge noninvariant couplings other than potential terms are not
included into the theory. For the theory in which gauge invariance is not
required from the outset this is in fact the simplest generalization of a
conventional SUSY QED. This superfield potential (\ref{pot1a}) is turned out
to be generically unstable unless SUSY is spontaneously broken. However, it
appears not to be enough. To provide an overall stability of the potential
one additionally needs the special direct constraint being put on the vector
superfield itself that is made by an appropriate SUSY invariant Lagrange
multiplier term (\ref{ext}). Remarkably enough, when this term is written in
field components it leads precisely to the nonlinear $\sigma $-model type
constraint of type (\ref{const}) which one has had in the non-SUSY case. So,
we come again to the picture, which we called the inactive SLIV, with a
Goldstone-like photon and special (SLIV restricted) gauge invariance
providing the cancellation mechanism for physical Lorentz violation. But now
this picture follows from the vacuum stability and supergauge invariance in
the extended SUSY QED rather than being postulated as is in the non-SUSY
case. This allows to think that a generic trigger for massless photons to
dynamically emerge happens to be spontaneously broken supersymmetry rather
than physically manifested Lorentz noninvariance.

In more exact terms, in the broken SUSY phase one eventually comes to the
almost standard SUSY QED Lagrangian (\ref{fin}) possessing some special
gauge invariance emerged. This invariance is only restricted by the gauge
condition put on the vector field, $A_{\mu }A^{\mu }=|S|^{2}$, which appears
to be invariant under supergauge transformations. One can use this
supergauge freedom to reduce the nondynamical scalar field $S$ to some
constant background value so as to eventually come to the nonlinear vector
field constraint (\ref{const}). As a result, the inactive time-like SLIV is
properly developed, thus leading to essentially nonlinear emergent SUSY QED
\ in which the physical photon arises as a three-dimensional Lorentzian NG
mode. So, figuratively speaking, the photon passes through three evolution
stages being initially the massive vector field component of a general
vector superfield (\ref{lag3}), then the three-level massless companion of
an emergent photino in the broken SUSY stage (\ref{der}) and finally a
generically massless state as an emergent Lorentzian mode in the inactive
SLIV stage (\ref{const1}).

As to an observational status of emergent SUSY theories, one can see that,
as in an ordinary QED, physical Lorentz invariance is still preserved in the
SUSY QED model at the renormalizable level and can only be violated if some
extra gauge noninvariant couplings (being supersymmetric analogs of the
high-dimension couplings (\ref{dim5})) are included into the theory.
However, one may have some specific observational evidence in favor of the
inactive SLIV even in the minimal (gauge invariant) supersymmetric QED and
Standard Model. Indeed, since as mentioned above the vacuum stability is
only possible in spontaneously broken SUSY case, this evidence is related to
an existence of an emergent goldstino-photino type state in the SUSY visible
sector. Being mixed with another goldstino appearing from a spontaneous SUSY
violation in the hidden sector this state largely turns into the light
pseudo-goldstino. Its study seem to be of special observational interest for
this class of models that, apart from some indication of an emergence nature
of QED and the Standard Model, may appreciably extend the scope of SUSY
breaking physics being actively studied in recent years. We may return to
this important issue elsewhere.

\section*{Acknowledgments}

I thank Colin Froggatt, Alan Kostelecky, Rabi Mohapatra and Holger Nielsen
for stimulating discussions and correspondence. Discussions with the
participants of the International Workshop \textquotedblright What Comes
Beyond the Standard Models?\textquotedblright\ (14--21 July 2013, Bled,
Slovenia) are also appreciated. This work was supported in part by the
Georgian National Science Foundation under grant No. 31/89.


\begin{thebibliography}{99}
\bibitem{NJL} Y.~Nambu, G.~Jona-Lasinio, Phys. Rev. 122 (1961) 345;

J. Goldstone, Nuovo Cimento 19 (1961) 154.

\bibitem{bjorken} J.D.~Bjorken, Ann. Phys. (N.Y.) \ 24 (1963) 174;

P.R. Phillips, Phys. Rev. 146 (1966) 966 ;

T.~Eguchi, Phys. Rev. D 14 (1976)\ 2755.

\bibitem{cfn} J.L.~Chkareuli, C.D.~Froggatt, H.B.~Nielsen,
Phys.~Rev.~Lett.~87 (2001)\ 091601, hep-ph/0106036; Nucl.~Phys.~B \ 609
(2001) 46, hep-ph/0103222.

\bibitem{jb} J.D. Bjorken, hep-th/0111196.

\bibitem{kraus} P. Kraus, E.T. Tomboulis, Phys. Rev. D 66 (2002) 045015.

\bibitem{jen} A. Jenkins, Phys. Rev. D 69 (2004) 105007.

\bibitem{bluhm} V.A. Kostelecky, Phys. Rev. D 69 (2004) 105009 ;

R.~Bluhm, V.~A.~Kostelecky, Phys.\ Rev.\ D 71 (2005) 065008.

\bibitem{alan} V.A. Kostelecky, S. Samuel, Phys. Rev. D \textbf{39 }(1989)
683;

V.A. Kostelecky, R. Potting, Nucl. Phys. B\textbf{\ 359 }(1991) 545.

\bibitem{ted} T. Jacobson, D. Mattingly, Phys. Rev. D 64, 024028 (2001).

\bibitem{refs} S. Chadha, H.B. Nielsen, Nucl. Phys. B 217 (1983) 125;

S.M. Carroll, G.B. Field, R. Jackiw, Phys. Rev. D 41 (1990) 1231.

\bibitem{2} D. Colladay, V.A. Kostelecky, Phys. Rev. D 55 (1997) 6760 ; D 58
(1998) 116002;

V.A. Kostelecky, R. Lehnert, Phys. Rev. D 63 (2001) 065008.

\bibitem{3} S. Coleman, S.L. Glashow, Phys. Lett. B 405 (1997) 249; Phys.
Rev. D 59 (1999) 116008.

\bibitem{nambu} Y. Nambu, Progr. Theor. Phys. Suppl. Extra  (1968) 190.

\bibitem{az} A.T. Azatov, J.L. Chkareuli, Phys. Rev. D 73 (2006) 065026;
ArXiv: hep-th/0511178.

\bibitem{kep} J.L. Chkareuli, Z.R. Kepuladze, Phys. Lett. B 644 (2007) 212;
hep-th/0610277.

\bibitem{jej} J.L. Chkareuli, J.G. Jejelava, Phys. Lett. B 659 (2008) 754;
ArXiv:0704.0553 [hep-th].

\bibitem{urr} O.J. Franca, R. Montemayor, L.F. Urrutia, Phys.Rev. D 85
(2012) 085008.

\bibitem{gra} J.L. Chkareuli, J.G. Jejelava, G. Tatishvili, Phys. Lett. B
696 (2011) 124; ArXiv:1008.3707 [hep-th].

\bibitem{GL} S. Weinberg, The Quantum Theory of Fields, v.2, Cambridge
University Press, 2000.

\bibitem{vru} R.~Bluhm, N.L.~Gagne, R.~Potting, A.~Vrublevskis, Phys.\ Rev.\
D 77 (2008) 125007.

\bibitem{par} J.L. Chkareuli, Z. Kepuladze, G. Tatishvili, Eur. Phys. J. C
55 (2008) 309, ArXiv:0802.1950 [hep-th]; J.L. Chkareuli, Z. Kepuladze, Eur.
Phys. J. C 72 (2012) 1954, ArXiv:1108.0399 [hep-th].

\bibitem{pos} S. Groot Nibbelink, M. Pospelov, Phys. Rev. Lett. 94, 081601
(2005).

\bibitem{pos1} P. A. Bolokhov, S. G. Nibbelink, M. Pospelov, Phys. Rev. D
72, 015013 (2005).

\bibitem{puj} O. Pujolas, S. Sibiryakov, JHEP 1201 (2012) 062.

\bibitem{j} J.L. Chkareuli, Phys. Lett. B 721 (2013) 146; ArXiv:1212.6939
[hep-th].

\bibitem{wess} H.P. Nilles, Phys. Rep. 110 (1984) 1;

J. Wess, J. Bagger, Supersymmetry and Supergravity, 2nd ed., Princeton
University Press, Princeton, 1992;

S.P. Martin, A Supersymmetry Primer, hep-ph/9709356.

\bibitem{ross} R. Hodgson, I. Jack, D.R.T. Jones, G.G. Ross, Nucl. Phys. B
728 (2005) 192.

\bibitem{vis} K. Izawa, Y. Nakai, T. Shimomura, JHEP 1103 (2011) 007.

\bibitem{tha} D. Bertolini, K. Rehermann, J. Thaler, JHEP 1204 (2012) 130.

\bibitem{gol} C. Cheung, Y. Nomura, J. Thaler, JHEP 1003 (2010) 073;

N. Craig, J. March-Russell, M. McCullough, JHEP 1010 (2010) 095;

H.-C. Cheng, W.-C. Huang, I. Low, A. Menon, JHEP 1103 (2011) 019;

R. Argurio, Z. Komargodski, A. Mariotti, Phys. Rev. Lett. 107 (2011) 061601.

\bibitem{h} ATLAS Collaboration, G. Aad et. al., Phys. Lett. B 716 (2012) 1;

CMS Collaboration, S. Chatrchyan et. al., Phys. Lett. B 716 (2012) 30.
\end{thebibliography}
\end{document}